\documentclass[12pt]{article}
\usepackage{graphicx,amssymb,epsfig,epsf} 
 \usepackage{ulem}
\usepackage[usenames]{color}
\usepackage{rotating}
\usepackage{epsfig}
\usepackage{amsmath}

\newcommand{\sbot}[1]{\tilde{b}_#1}
\newcommand{\sstop}[1]{\tilde{t}_#1}

\newcommand{\sbotc}[1]{\tilde{b}_#1^*}
\newcommand{\sstopc}[1]{\tilde{t}_#1^*}

\def\ptmiss{\not\!\!{p_T}}

\begin{document}

\begin{flushright}
   {\bf KIAS-P10045}\\
   {\bf TIFR/TH/10-36}\\ 
\end{flushright}

\vskip 30pt

\begin{center}
{\large \bf Boosted top quarks in supersymmetric cascade decays at the LHC}\\
\vskip 20pt
{Priyotosh Bandyopadhyay$^{a}$\footnote{priyotosh@kias.re.kr}, {Biplob Bhattacherjee$^{b}$\footnote{biplob@theory.tifr.res.in}}}  \\
\vskip 20pt
{$^a$ Korea Institute for Advanced Study, \\
 Hoegiro 87(207-43 Cheongnyangni-dong), \\
 Seoul 130-722, Korea}\\
\vskip 20pt
{ $^{b}$ Department of Theoretical Physics,\\ Tata Institute of Fundamental Research, \\ 
1, Homi Bhabha Road, Mumbai 400 005, India. \\}

\end{center}

\vskip 65pt

\abstract{ At the LHC, a generic supersymmetric cascade can be a source of 
 top quark. Specifically third generation squarks and gluino are the major 
sources of top quark which could also be boosted. In this article, we have 
shown that jet substructure algorithm can be very useful in identifying such boosted 
top quarks in the cascade. We take inclusive three jets plus zero lepton plus 
missing energy final state and try to reconstruct at least one hadronically 
decaying top quark by using top tagging technique which has good prospect at 
the LHC. }\\

\renewcommand{\thesection}{\Roman{section}} 
\setcounter{footnote}{0} 
\renewcommand{\thefootnote}{\arabic{footnote}} 
\newpage
\section{Introduction}
Standard Model (SM) has been successful in explaining a wide variety of phenomena and most of the 
experimental results are consistent with the SM \cite{Langacker:2009my}.  However, there are some issues that make it incomplete as a theory. One of the most important issues is the mass of the 
Higgs boson, the particle responsible for giving mass to the fermions and the gauge bosons, is 
itself radiatively unstable. There is no solution to this puzzle in the SM and hence one feels justified to go beyond the SM and 
look for new physics. Supersymmetry (SUSY) \cite{SUSY} is one of the possible extension of SM which can protect Higgs mass and in principle solve the problem. Supersymmetry predicts superpartners 
 of SM particles. Since no super partner has been observed so far, they must be heavier than 
the corresponding experimental limits \cite{Amsler:2008zzb}. This simply implies that the 
supersymmetry must be broken. Unfortunately, the SUSY breaking mechanism is completely unknown. 
For this reason, low energy SUSY spectrum can be completely arbitrary and it needs more than 100 parameters to specify the masses and couplings of the SUSY particles. Practically it is very difficult to explore such a multi-dimensional parameter space and study its collider signatures \cite{SUSYSCAN}. Therefore, it is necessary to adopt some specific assumptions for SUSY breaking mechanisms. There are several phenomenologically viable SUSY breaking  mechanisms. Gravity mediated supersymmetry breaking model like constrained MSSM (cMSSM) \cite{mSUGRA} is the most popular amongst all. In cMSSM, the full spectrum is completely determined only by four and half parameters 
specified at the high scale.  These are universal scalar mass ($m_0$), universal gaugino mass 
($m_{1/2}$), tri-liner coupling constant ($A_0$), the ratio between the vacuum 
expectation values of up and down type Higgs fields ($\tan{\beta}$) and the sign of 
Higgs mixing term 
($\mu$) in the super-potential. This small set of parameters makes the model very predictive. \\

In a R parity\footnote{R parity is a discrete symmetry defined as a $R_p$=$(-1)^{(3B+2S+L)}$, 
where L, B and S are lepton number, baryon number and spin of the particle respectively.} conserving supersymmetric model, sparticles must be produced in pair and eventually cascade down to the lightest SUSY particle (LSP) which can not decay to SM particles. Now, if neutralino is the LSP, 
 because of its weakly interacting nature it could be a good dark matter candidate. In a 
collider experiment it is impossible to detect such weakly interacting particle, thus 
contributes as missing energy in the final state. In general, the pair production of SUSY particles will lead to final states with multiple jets and leptons plus missing energy. So far, many 
detailed studies have been carried out to discover SUSY at the LHC \cite{SUSYSEARCH}, particularly in the context of cMSSM. Such studies indicate that inclusive jets plus missing energy channel with 
zero lepton has the highest reach in the cMSSM parameter space \cite{zerolepton}. The final state with isolated leptons are also important, since the requirement of one or more isolated leptons 
can suppress huge QCD background, although such channels have lower reach than the zero lepton channels \cite{onelepton}.\\

If low scale supersymmetry exists in nature, the study of third generation squarks at the LHC is of 
a special interest. This is because, the third generation Yukawa coupling is relatively larger than the rest,  
which results in  relatively larger mass splitting between the mass eigenstates. Thus the lighter eigenstates could become light enough and might be discovered at the Tevatron or in the early stage of LHC \cite{stop_sbottom}. \\

Third generation squarks ($\tilde{t}_1$, $\tilde{t}_2$ and $\tilde{b}_1$, $\tilde{b}_2$) can  be pair produced at the LHC and these can be produced from the decay of SUSY particles. 
Depending on the parameter space, the dominant decay modes of stop squarks 
($\tilde{t}_1$, $\tilde{t}_2$) are  $ t~\tilde{\chi}_{1,2,3,4}^0$ and  $b~\tilde{\chi}_{1,2}^{\pm}$. 
Similarly sbottom 
squarks ($\tilde{b}_1$, $\tilde{b}_2$) can decay to $ t ~\tilde{\chi}_{1,2}^{\pm}$ and 
$ b~\tilde{\chi}_{1,2,3,4}^0$. If the 
squarks are heavier than gluino except the third generation, gluino will decay entirely to the third 
generation squarks. If all squarks are heavier than gluino, the gluino will decay to three body final state 
through off shell squarks. In general, stop and sbottom squarks, being lighter than the first two generation 
squarks, contribute more to the gluino three body decay. This is why, final states with top and bottom quarks in the gluino three body decay are relatively more favoured. Final states with $b$ jets accompanying 
with multiple jets, leptons and missing energy have been studied extensively. Another interesting possibility is 
top rich final states from gluino decay. It is thus important to identify the presence of top quark 
in the SUSY cascade, as it carries information about third generation squarks. 
However, top quark identification in a SUSY cascade is not an easy task. Top quark mostly decays 
to W and b quark unless charged Higgs boson is lighter than top quark. In case of leptonic decay 
of W, the neutrino, that is present in the final state also contributes to the missing energy, 
in addition to the missing energy from the lightest neutralino. In order to reconstruct semileptonically decaying top, we have to separate out the neutrino missing energy contribution 
coming from the top quark. This is why, it is difficult to reconstruct a top quark by its semileptonic decay in the cascade. On the other hand, hadronically decaying top quark has no real missing energy contribution. However, it is very difficult to choose the correct combination of three jets from multiple jets (which are always present in a typical SUSY cascade) as it suffers from combinatorial backgrounds. It is thus difficult, although, not impossible to identify top quarks in a SUSY cascade. Beside this, top quarks in SUSY cascade can be highly boosted due to the usual separation between strong and electroweak sector of MSSM. Because of the high boost, the decay products of top quark are generally collimated. It makes it very difficult to isolate 
the decay products as separate jets or leptons; and often ends up with a single fat jet in the final state. \\

From the above discussion we have seen that the identification of top quark in SUSY cascade faces two types of difficulties, e.g., combinatorial background in top reconstruction and collimation of the decay products of the   
highly boosted top quarks. Recently, a new method has been proposed to distinguish jets originated from 
boosted heavy particles by the method of jet substructure. By using this method, one can fully reconstruct 
the decaying heavy particle through its hadronic final state. A number of recent studies have used this 
technique to identify boosted top quarks \cite{Agashe:2006hk,boostedtop,kaplan}, Higgs bosons \cite{boostedhiggs}, or $W/Z$ bosons \cite{boostedw} and the results are very encouraging. It is therefore, a good idea to use the above mentioned technique to reconstruct the top signals in the SUSY cascade.\\

In this paper we systematically study the top quark production in the SUSY cascade and its identification 
by newly developed top tagging method at the LHC with centre of mass energy 14 TeV. We take inclusive three jets plus zero lepton plus missing energy final state and try to reconstruct at least one hadronically decaying top quark by using top tagging technique, described in \cite{kaplan}.  
Of course, the production of the boosted top will depend on the mass of the 
parent squark (stop and sbottom), or gluino and their corresponding branching
fraction to the top quark. In ref~\cite{Plehn} authors have studied the pair production of lightest stop squarks 
assuming the branching $\tilde{t}_1 \rightarrow t \tilde{\chi}_1^0$ to be equal to 1 and they have 
shown that it is possible to extract the stop ($\tilde{t}_1$) - neutralino ($\tilde{\chi}_1^0$) mass difference.
 However, this assumption does not hold as we change the parameter space. \\

The article is organized as follows. In section II we describe the origin of top quark final states from the 
decay of third generation squarks and gluino and also discuss the effective branching to top quark for 
different scenarios. In section III, we define our benchmark points for the analysis. We discuss our results,
 briefly addressing the top tagging technique in section IV. Finally we conclude in section V 
 discussing the possible issues of this work. \\

\section{Possible Scenarios}
In this section we shall describe the possible sources of top quark from the decay of sparticles. In general, 
the squark mass spectra are very different as one goes from first two generations to third generation. The off 
diagonal term in the squark mass matrix is given by  $m_f (A_f + \mu R_f)$ where $R_f=\cot \beta$ for up type 
squark and  $R_f=\tan \beta$ for down type squark and $m_f$ is the corresponding quark mass 
\cite{thirdgeneration}. As first two generation 
quark masses are negligible, we can neglect the off diagonal terms in the mass matrix. The general hypothesis of 
flavour blind soft parameter for first two generations avoid potentially dangerous FCNC and CP-violating effects in MSSM. However, masses of the third generation squarks are more complicated because of the Yukawa couplings and the corresponding changes in RG equations. Due to large mixing through the off diagonal terms, there could be large splittings between the mass eigenvalues. Thus, in general the third generation masses are quite non-degenerate and lighter ones ($\tilde{t}_1$, 
$\tilde{b}_1$) can become rather light. \\
 
We are now ready to discuss the possible decay modes of third generation squarks. If third generation squarks are heavier than gluino, preferably decay to corresponding quark and gluino, if kinematically  allowed. In case, gluino 
is heavier than stop or sbottom squark, there are several possible interesting channels like
\begin{center}
$\tilde{t}_{1,2} \rightarrow t \tilde{\chi}_{i}^0$ (i=1-4) ~~~~ $\tilde{t}_{1,2} \rightarrow b \tilde{\chi}_{i}^{\pm}$~~ (i=1-2)\\
~~$\tilde{b}_{1,2} \rightarrow b \tilde{\chi}_{i}^0$ (i=1-4) ~~~~ $\tilde{b}_{1,2} \rightarrow t \tilde{\chi}_{i}^{\pm}$~~ (i=1-2) .\\
\end{center}

If the mass difference between $\tilde{t}_2$ ($\tilde{b}_2$) and $\tilde{t}_1$ ($\tilde{b}_1$)  or 
$\tilde{t}_2$ ($\tilde{b}_2$) and $\tilde{b}_1$ ($\tilde{t}_1$) is large enough, the following decays \cite{Bartl:1998xk}

\begin{center}
$\tilde{t}_{2} \rightarrow \tilde{t}_{1} ~~Z,/H,/A $  ~~~~~~ $\tilde{t}_{2} \rightarrow \tilde{b}_{1} ~~~W^{+}/H^{+}$ \\
$\tilde{b}_{2} \rightarrow \tilde{b}_{1} ~~Z,/H,/A $  ~~~~~~ $\tilde{b}_{2} \rightarrow \tilde{t}_{1} ~~~W^{+}/H^{+}$ \\
\end{center}
are also possible. Several studies have been performed in this context of CP conserving \cite{cpc_cascade} and 
CP violating MSSM \cite{cpv_cascade}. 
For a wide range of parameter space, all tree level two body decay modes of lightest stop 
squark are forbidden and the loop decay $\tilde{t}_1 \rightarrow c \tilde{\chi}_1^{0}$ becomes significant 
\cite{stoploop}. In the case where $\tilde{t}_{1}$ is the NLSP, it decays to three body \cite{Porod:1996at} or 
four body final state \cite{stop4body}. 
In the next few paragraphs, we discuss the effective decay branching fractions to top quark from squark and gluino 
including third generation. We use SUSY-HIT \cite{SUSY-HIT} for the calculation of decay branching fractions in our 
analysis.\\

\begin{figure}[t]
\begin{center}
\hskip -15pt
{\epsfig{file=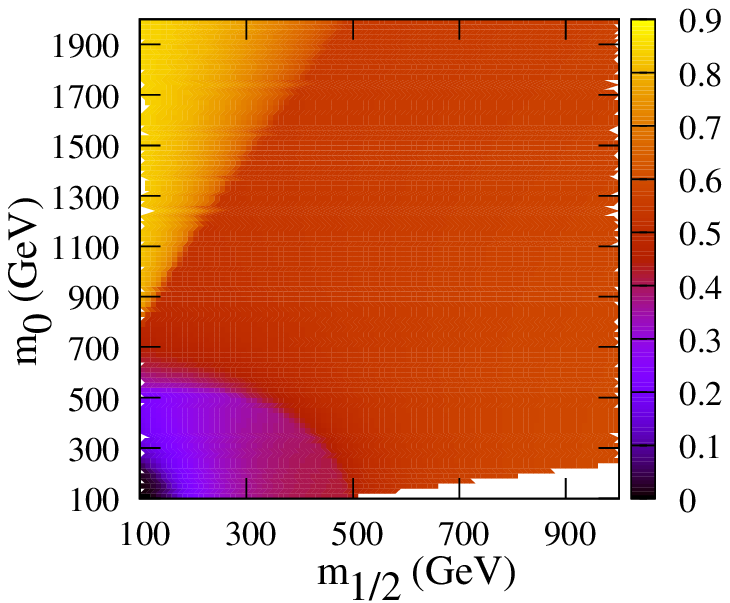,width=8.0 cm,height=8cm,angle=0}}
\hskip 12pt 
{\epsfig{file=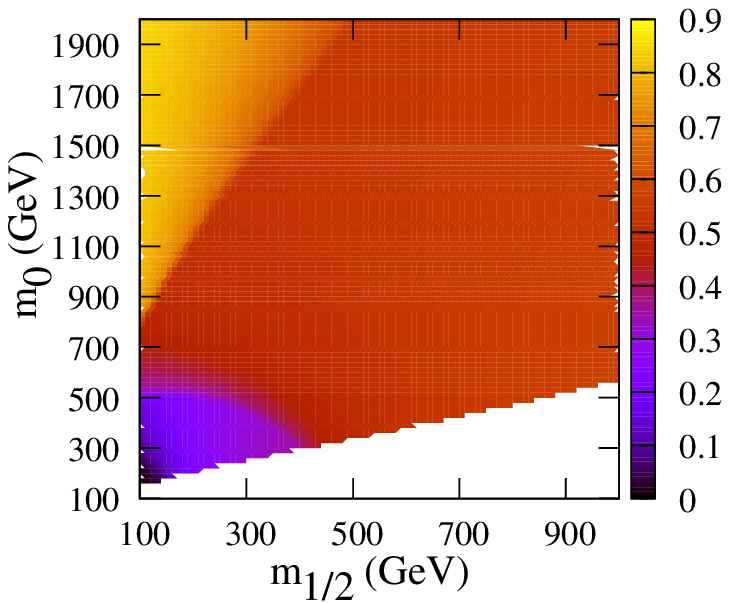,width=8.0cm,height=8cm,angle=0}}
\caption{\sf Branching fraction of $\tilde{t}_1 \to t \tilde{\chi}_i /\tilde{g}$ for $\tan\beta$=10 (left) and 
$\tan\beta$=50 (right) in $m_{0} - m_{1/2}$ plane for $A_0=0$, and $\mu$ $>$0.}\label{stop}
\end{center}
\end{figure}

\begin{figure}[t]
\begin{center}
\hskip -15pt
{\epsfig{file=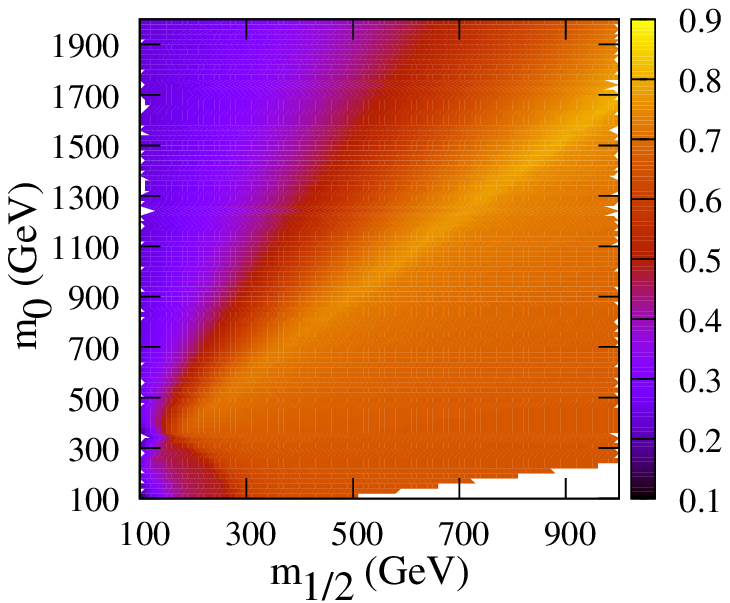,width=8.0 cm,height=8cm,angle=0}}
\hskip 12pt 
{\epsfig{file=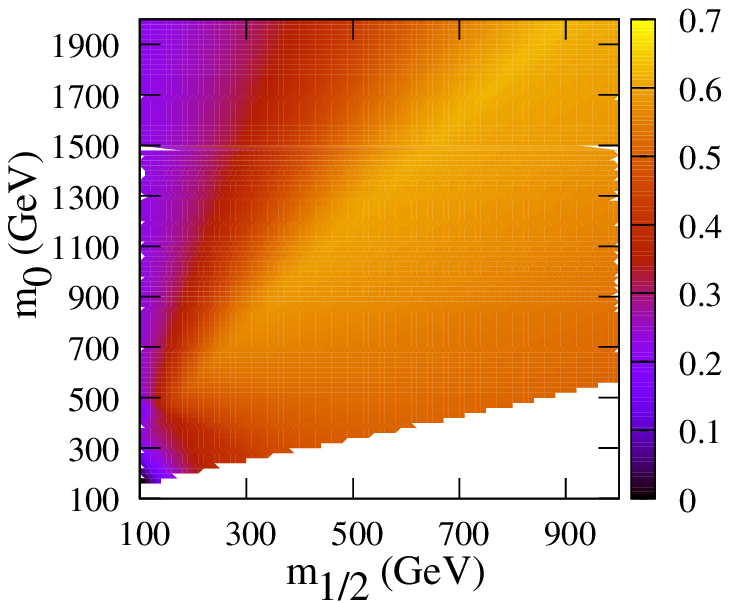,width=8.0cm,height=8cm,angle=0}}
\caption{\sf Branching fraction of $\tilde{b}_1 \to t \tilde{\chi}_i^{-}$ for $\tan\beta$=10 (left) and 
$\tan\beta$=50 (right) in $m_{0} - m_{1/2}$ plane for  $A_0=0$, and $\mu$ $>$0.}\label{sbottom}
\end{center}
\end{figure}

\noindent {\bf $\tilde{t}_1$, $\tilde{t}_2$ to top quark}: We have already mentioned the possible decay modes of 
$\tilde{t}_1$ and $\tilde{t}_2$ squarks. We expect that the top quark contribution from $\tilde{t}_1$  will be more 
than $\tilde{t}_2$, because of higher mass value of later. In most of the cases $\tilde{t}_1$ is much lighter 
than $\tilde{t}_2$ and thus $\tilde{t}_1$ pair production cross section is significant.  
In order to understand the relevant branching of lightest stop squark to top quark, we scan over $m_0$-$m_{1/2}$ plane of the cMSSM parameter space for $\tan \beta$ =10 and 50, $A_0$ =0 and $\mu>0$. We vary $m_0$ from 100 to 2000 GeV 
and $m_{1/2}$ from 100 to 1000 GeV. In Figure \ref{stop}, we show the $\tilde{t}_1$ to top branching ratio in the $m_0$ - $m_{1/2}$ plane. Here we 
have included $\tilde{t}_1 \rightarrow t \tilde{\chi}_i^0$ and  $\tilde{t}_1 \rightarrow t \tilde{g}$ 
branching fractions. The white region 
is excluded because $\tilde{\tau}_1$ is the LSP in that region. We have seen that for small $m_0$ and $m_{1/2}$,  $\tilde{t}_1$ to top quark branching is small ($\sim$ 0.2 to 0.3) for both values of $\tan{\beta}$. In this region, $\tilde{t}_1$ is light and $\tilde{t}_1 \rightarrow t \tilde{\chi}_{i}^0$ is phase space suppressed and $\tilde{t}_1 \rightarrow b \tilde{\chi}_i^{\pm}$ is dominant. The another reason for it is that, in cMSSM, $\tilde{\chi}_1^{0}$ is often bino like and $\tilde{\chi}_1^{\pm}$ is wino like and bino-squark-quark coupling is small compared to wino-quark-squark coupling. For large value of $m_{0}$ and small $m_{1/2}$,  $\tilde{t}_1$ is heavier than gluino and $\tilde{t}_1 \rightarrow t \tilde{g}$ is dominant due to the strong coupling 
(see yellow region of Figure \ref{stop}).  But in this case, stop squarks are very heavy and production cross section is negligibly small. For considerably large value of stop mass, $m_{\tilde{t}_1}$ is larger than   $m_{\tilde{\chi}_{2}^0} + m_t$. Since $\tilde{\chi}_2^0$ is wino like in 
most of the cases in cMSSM model, $\tilde{t}_1 \rightarrow \tilde{\chi}_{2}^0~t$ is significant (see red region of Figure \ref{stop}). In the similar way, $\tilde{t}_2$ squark can also produce top quarks, but, such effects are in general negligible due to small cross section. \\

\noindent {\bf $\tilde{b}_1$, $\tilde{b}_2$  to top quark}: In Figure \ref{sbottom}, we show the branching of $\tilde{b}_1 \rightarrow t \tilde{\chi}_{1,2}^{\pm}$ in the $m_{0}$ - $m_{1/2}$ plane for $\tan{\beta}=$10 and 50. We can see from the figure that, where $\tilde{t}_1 \to t \tilde{\chi}_i$ branching fraction is large, the branching $\tilde{b}_1 \to t \tilde{\chi}_{1}^{\pm}$ is small. This behaviour remains the same going  from $\tan{\beta}=10$ to 50, though for the later case it shrinks a bit. For large $m_0$ and small $m_{1/2}$, higgsino component in the chargino 
is enhanced because in this region $\mu$ becomes small compared to bino ($M_1$) and wino ($M_2$) soft mass parameters. 
Otherwise as mentioned earlier that 
$\tilde{\chi}_{1}^{\pm}$ is wino like in most of the cMSSM parameter space and for this 
reason $\tilde{b}_1 \rightarrow t \tilde{\chi}_{1}^{\pm}$ is favoured rather than 
$\tilde{b}_1 \rightarrow b \tilde{\chi}_{1}^{0}$. This complementary behaviour adds to an extra 
motivation for the search of boosted top quarks in the SUSY cascade in which $\tilde{b}_1$ is present. 
In general, sbottom squarks are heavier than stop squarks and thus, the contributions to top quark production 
may be small. Similarly, $\tilde{t}_2$, $\tilde{b}_2$  also contribute to the top quark final state. \\

\begin{figure}[t]
\begin{center}
\hskip -15pt
{\epsfig{file=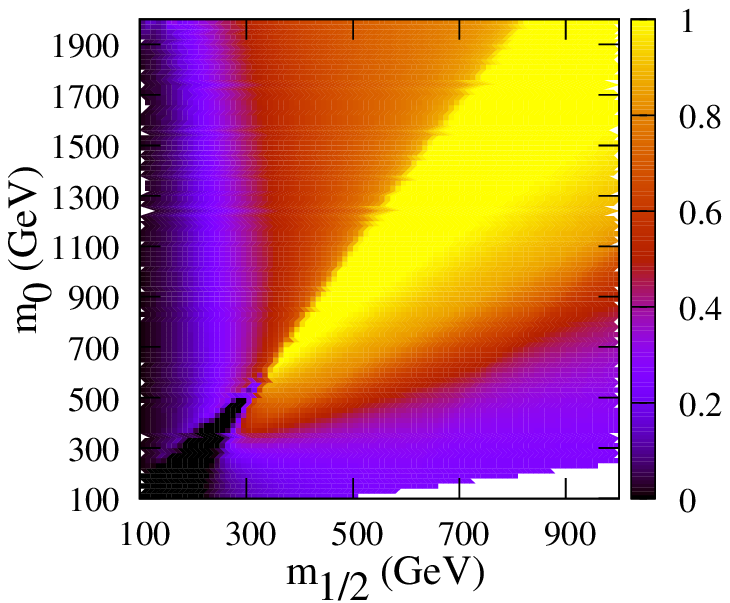,width=8.0 cm,height=8cm,angle=-0}}
\hskip  12pt 
{\epsfig{file=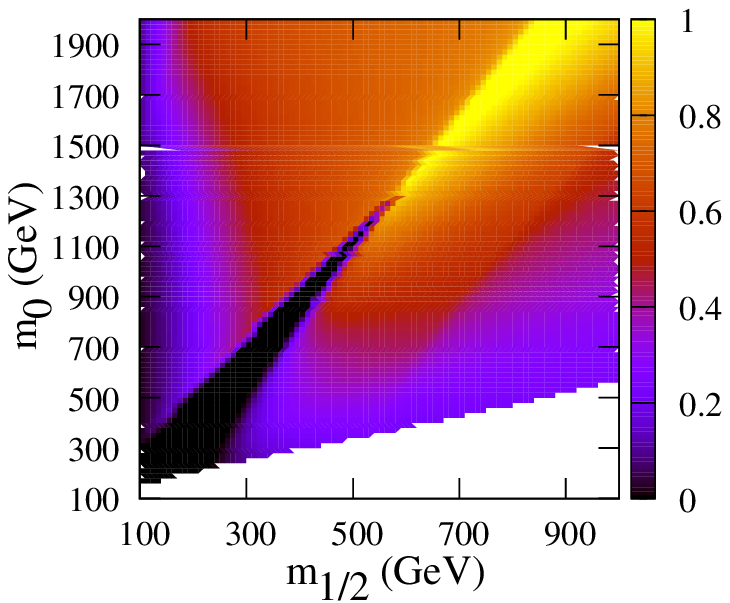,width=8.0cm,height=8cm,angle=-0}}
\caption{\sf Branching fraction of $\tilde{g}$ to top quark via two body and three body channels 
for $\tan\beta$=10 (left) and $\tan\beta$=50 (right) in the $m_{0} - m_{1/2}$ plane for $A_0=0$ and 
$\mu>0$.}\label{gluinotop}
\end{center}
\end{figure}

\begin{figure}[t]
\begin{center}
\hskip -15pt
{\epsfig{file=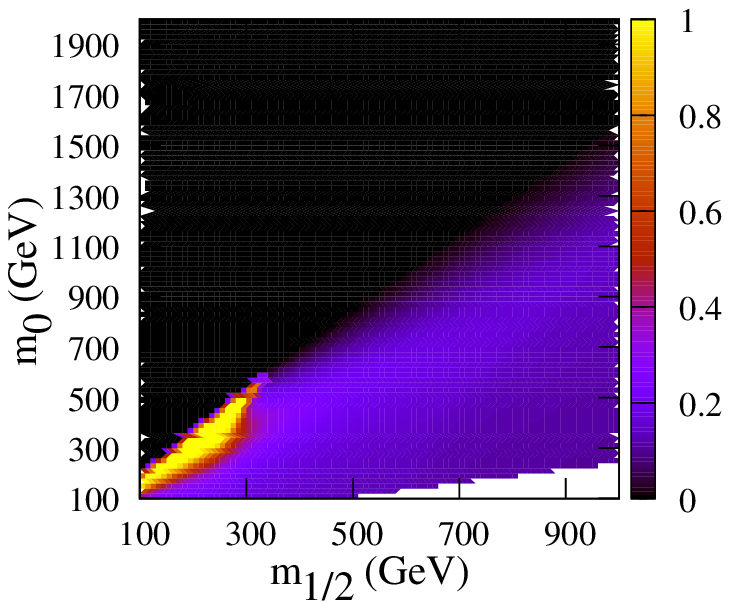,width=8.0 cm,height=8cm,angle=-0}}
\hskip  12pt 
{\epsfig{file=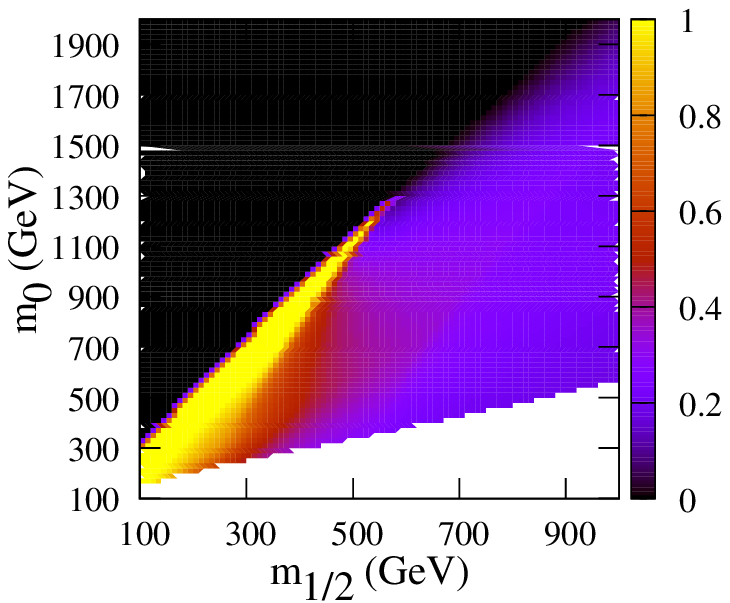,width=8.0cm,height=8cm,angle=-0}}
\caption{\sf Branching fraction of $\tilde{g} \to (\bar{b} \sbot1 \,+\,c.c)$ 
for $\tan\beta$=10 (left) and $\tan\beta$=50 (right) in the $m_{0} - m_{1/2}$ plane for $A_0=0$ and $\mu>0$.}\label{gluinosbottom}
\end{center}
\end{figure}

\noindent {\bf $\tilde{g}$ contribution to top quark:} If the  gluino mass is above the third 
generation scalar quark mass, gluino decays through $\tilde{g} \to q \tilde{q}^*$. 
The corresponding branching fractions to different flavour modes 
depend on the associated mass values of the quarks and squarks. 
Otherwise if gluino is lighter than squarks, it can still decay 
to top quark via three-body decay \cite{gluinodecay}. Figure \ref{gluinotop} describes
the gluino to top branching fraction via two-body and three-body decays
 in cMSSM scenario for $\tan\beta$=10 and 50. There are fair amount of regions where 
Br($\tilde{g} \to t \sstopc1$) can be very high, whereas, Br($\tilde{g} \to b \sbotc1$) in such 
regions is very low (see Figure \ref{gluinosbottom}). There is another
 interesting region in Figure \ref{gluinosbottom} (see conical region) where $\tilde{g}$ 
almost decays to $b \sbotc1$. This is because, in this region, both stop and sbottom squarks are lighter 
than gluino, but, $m_{\tilde{g}} - m_{\sstop1}$ is less than SM top quark mass. Here the only allowed two body 
decay mode of gluino is $\tilde{b}_1 b$. The size of such region increases for large value of $\tan\beta$ \cite {Bartl:1994bu}.\\

If the gluino mass is lower than the stop and sbottom masses, the gluino decays through three body channel 
to $q \bar{q} \tilde{\chi}_i^0$ or $q \bar{q}^{'} \tilde{\chi}_{i}^{\pm}$. The $\sstop1$ or $\sbot1$  squarks are generally lighter than other squarks and these contribute to the mode  $t \bar{t} \tilde{\chi}_i^0$ and $t \bar{b} \tilde{\chi}_{i}^{\pm}$. A region of this kind is possible for higher values of $m_0$ relative to 
$m_{1/2}$. In the left black region gluino three body decay to top quark is kinematically disallowed. Gluino three body decay to top quarks may play a major role in the discovery of focus point region of cMSSM, in which only 
gauginos are light and squarks can be very heavy \cite{focuspoint}. \\

\noindent {\bf $\tilde{q}$ contribution to top quark:} If squarks are heavier than gluino, these can decay 
to gluino and eventually, gluino decays through three body channels discussed before. In this case, the 
gluino three body final states may contain top quarks. The other possibility is that, the gluino can be 
lighter than first two generations, but, it may be heavier than the third generation squarks. 
In that case, the decay chain like  $\tilde{q} \to \tilde{g} q$~$\to \sstop1 t q$  is possible.   

\begin{figure}[t]
\begin{center}
\hskip -15pt
{\epsfig{file=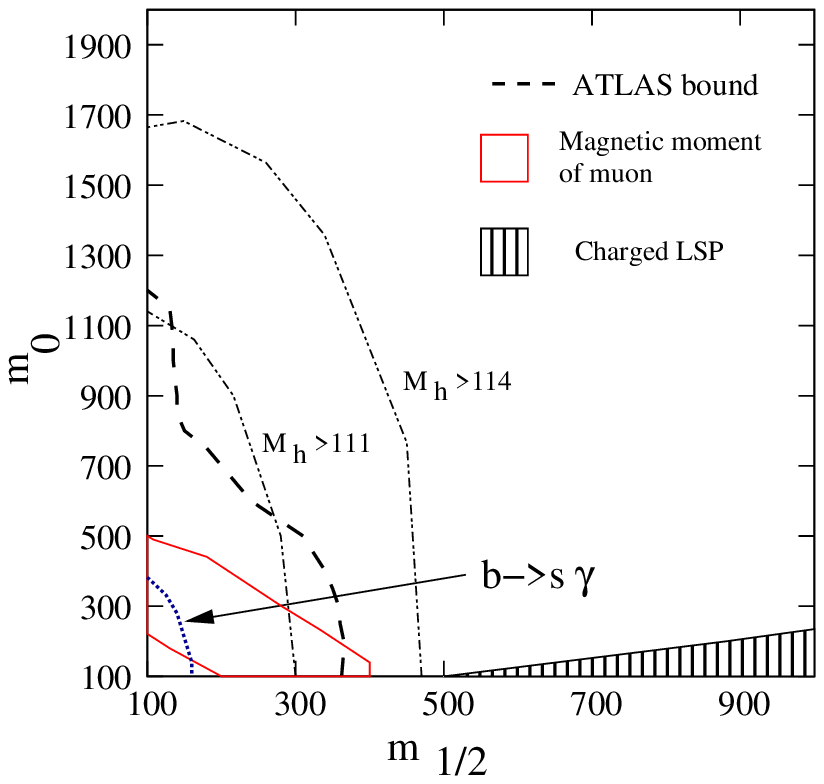,width=8.0 cm,height=8cm,angle=-0}}
\hskip  12pt 
{\epsfig{file=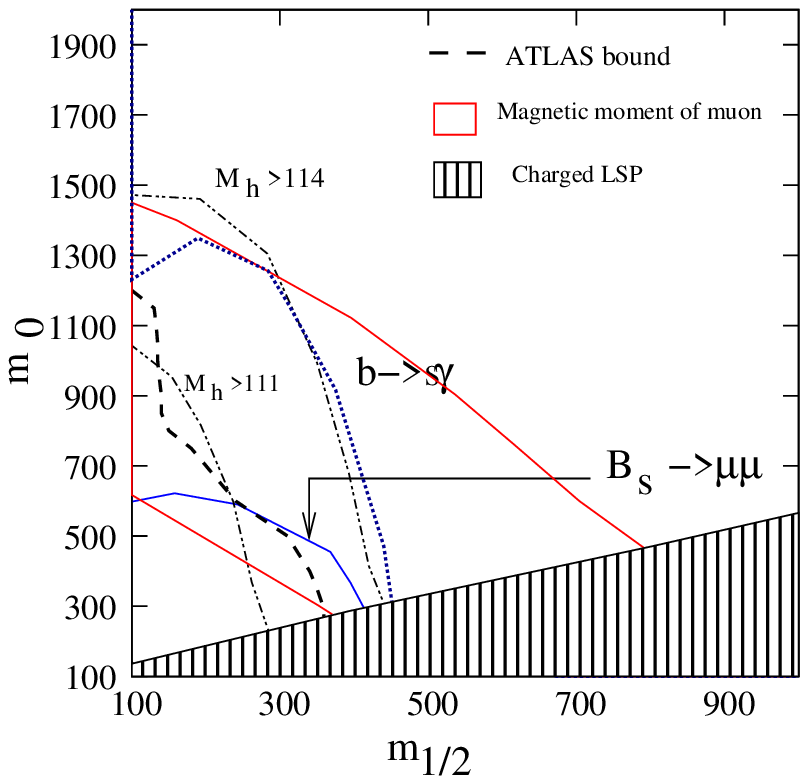,width=8.0cm,height=8cm,angle=-0}}
\caption{\sf Low energy and collider bounds in $m_0$-$m_{1/2}$ plane for $\tan\beta$=10, and $\tan\beta$=50. 
 }\label{lowenergy}
\end{center}
\end{figure}

\section{Benchmark Points}
In the previous section, we have discussed the possible sources of top quarks in the SUSY cascade in the context of cMSSM. Now, we explore such possibilities by choosing some 
benchmark points in the $m_0 -m_{1/2}$ plane. The cMSSM model is constrained by different low energy observables 
as well as direct searches like LEP, Tevatron and LHC. We have considered such bounds while defining the benchmark 
points. We scan over $m_0$ -$m_{1/2}$ plane for two fixed values of $\tan\beta$: 10 and 50 with sign of 
$\mu>$ 0 and $A_0$ =0. We take low energy bounds such as $b \rightarrow s \gamma$, 
$B_s \rightarrow \mu^{+} \mu^{-}$ and anomalous magnetic moment of muon. We closely follow the reference 
\cite{bhattacherjee} in our calculation and we use SuperIso \cite{superiso} software for this.  
We also consider recent bounds from 7 TeV LHC data with an integrated luminosity 
35 pb$^{-1}$ on cMSSM parameter space in the inclusive multi-jet plus missing energy channel. The current experimental lower bound on SM Higgs mass from LEP is 114.4 GeV and this bound is applicable to supersymmetric 
lightest Higgs, except for some specific regions in the SUSY parameter space. However, there is an estimated 
theoretical uncertainty of about 3 GeV on the Higgs mass due to higher order effects \cite{Heinemeyer}. For this 
reason we draw separate contours for $M_h$ =111 GeV and 114 GeV. In Figure \ref{lowenergy} we have shown corresponding bounds discussed above in the $m_0$ -$m_{1/2}$ for $\tan\beta$ 10 and 50. 
The region bounded by red box is favoured by magnetic moment of muon. The shaded region is excluded because 
in this region $\tilde{\tau_1}$ is the LSP. We have not shown $B_s \rightarrow \mu^{+} \mu^{-}$ bound for 
$\tan\beta$=10, because it is much weaker than other bounds considered here.

We shall tag the top quarks using jet 
substructure algorithm. This technique can only be applicable for highly boosted top 
quarks. The boost of top quarks in the SUSY cascade will depend on the relative 
separation between decaying particle, e.g., gluino, stop or sbottom squark and the 
decay products like neutralino and chargino. The mass difference between 
squark-gluino sector and electroweak gaugino sector is generally large in the 
cMSSM model. To cover different possibilities, we have chosen three specific 
benchmark points for further analysis. \\

We choose the following low mass point, which is just above the current LHC bound 

$$m_0=600 \, \rm{GeV}\quad \,\quad  m_{1/2}=350 \, \rm{GeV}\quad \,\quad A_0=0 \, \rm{GeV}\, \quad \mu>0 \,\quad and \quad \tan{\beta}=10 .$$

For this point the masses of $\sstop1$, $\tilde{\chi}_2^0$ and $\tilde{\chi}_1^0$ 
are 675 GeV, 267 GeV and  142 GeV respectively \footnote{For our analysis the 
top is taken to be 173 GeV.}. This ensures enough boost to the top, specially 
the one coming from $\sstop1 \to t \tilde{\chi}_1^0$. In this point 
Br($\sstop1 \to t \tilde{\chi}_1^0$)=18\% and 
Br($\sstop1 \to t \tilde{\chi}_2^0$)=10\%.
This implies that about $28\%$ of the stop squark will decay to a top quark. 
The stop squark mass is in the higher region and the pair production cross 
section is 93 fb which is much smaller than the total SUSY cross section 
of this point (about 5pb). We also take $\sbot1\sbotc1$ contribution for the 
analysis, which is about 22 fb, as the parameter point results in slightly 
heavier sbottom ($m_{\sbot1}=846$ GeV). The branching fraction of $\sbot1$ to 
the top and chargino mode is $\sim 70\%$ (from Figure \ref{sbottom}a). The chargino in 
this case can decay to a $W^\pm$ and a neutralino and the former can be a  
source of additional jets or leptons.\\

Interestingly, there is another mode which could also give rise to top quark.
This is  $\sbot1 \to \sstop1 W^-$, where the $\sbot1$ first decay to $\sstop1$ 
and the $\sstop1$ decays as usual. As Br($\sbot1 \to \sstop1 W^-)\sim 8\%$, 
contribution to  top, neutralino and $W^\pm$ final state is really small. \\

Gluino pair production cross section for this point is $\sim 709$ fb for $m_{\tilde{g}}= 859$ GeV, 
which adds to the third generation squarks through its decay, with an extra possible boosted top quark. 
The dominant decay mode of gluino is $\tilde{g} \to t \sstopc1$ with 
Br($\tilde{g} \to t \sstopc1) \sim 95$\% (see Figure \ref{gluinotop}). The squark gluino 
production is about 2.3 pb and it has the major contribution to the top 
final state through the gluino decay to top quark. \\ 

Next, we consider the contribution of gluino decay to sbottom squark. For this 
purpose, we choose the following parameter point from Figure \ref{gluinosbottom} as 
benchmark point 2 

$$m_0=900 \, \rm{GeV}\quad \,\quad  m_{1/2}=400 \, \rm{GeV}\quad \,\quad A_0=0 \, \rm{GeV} \quad \mu>0 \ \quad and \quad \tan{\beta}=50~ $$
where Br($\tilde{g} \to b \sbotc1 +c.c)=100\% $. The lightest sbottom squark produced from 
gluino further decays to top quark and chargino with branching fraction $\sim 59\%$. This 
choice leads us with $m_{\tilde{g}}=985$ GeV and $m_{\sbot1}=961$ GeV. 
For this benchmark point, lightest stop squark mass is about 840 GeV which gives no room 
for gluino decays to top quark. \\ 

So far, we have chosen benchmark points which are consistent with low energy constraints and 
direct search bounds. Now, we consider benchmark point 3 which satisfies the dark matter relic 
density constraint as well as other bounds
$$m_0=540 \, \rm{GeV}\quad \,\quad  m_{1/2}=490 \, \rm{GeV}\quad \,\quad A_0=0 \, \rm{GeV}\quad \quad \mu>0 \quad and \quad \tan{\beta}=50~.$$ \\

The mass of the lightest neutralino is 203 GeV and heavier Higgs masses are 
about 480 GeV. The decay widths of heavy Higgs bosons are large and these 
are about 15-20 GeV. Neutralinos can annihilate through heavy 
Higgs mediating s-channel processes and satisfy current bound of relic density.
We compute the relic density of this point by using micrOMEGAs 
\cite{micromegas} and we find that $\Omega h^2$ is 0.107 for top quark 
mass 173 GeV which is consistent with WMAP bound $\Omega h^2= 0.112 \pm 0.007$ at 95\% CL
\cite{WMAP}. \\

In this point left handed first two generation squarks are slightly heavier 
than gluino but right handed squarks are lighter. The gluino mass is about 1146 GeV.
The branching fractions of gluino to stop and sbottom squarks are $36\%$ and $21\%$ 
respectively. First two generation left handed squarks dominantly decay to 
electroweak gauginos, so, their branching to gluino is very small (about $2-3\%$).

\section{Analysis}

The identification of boosted top quarks has been described in the literature. Here we 
briefly describe our algorithm to identify boosted top quarks in the SUSY cascade. We have used 
PYTHIA\cite{PYTHIA} event generator for generating the events and hadronizations. 
We use FASTJET \cite{FASTJET} package for jet formation instead of default subroutine PYCELL in PYTHIA. 
The mass spectrum and decay branching fractions have been generated by SUSY-HIT. Here 
we define set of cuts used in our analysis.  

\begin{itemize}
\item We select events with at least 3 jets. The jets are formed by using Cambridge Aachen algorithm with a fixed R parameter to be equal to 1 \cite{cambridge_aachen}. One may vary this parameter to optimize the signal significance. The leading jet $p_T$ must be greater than 300 GeV and other two sub leading jets should have $p_T$  greater than 150 and 100 GeV respectively. Note that, $p_T$ of the jets are greater than the quoted value used by CMS collaboration for 
SUSY search in inclusive multi-jet plus missing energy channels \cite{zerolepton}. The absolute value of 
pseudorapidity of the leading jet must be less than 1.7. 
 
\item The missing $p_T$ must be greater than 300 GeV.  

\item We put veto on events with isolated leptons with $p_T$ greater than 10 GeV. 

\item  We use variable $R_1$ and $R_2$ which are given by 
$R_1 = \sqrt{\delta\phi_2^2 + (\pi -\delta\phi_1)^2}$ and 
$R_2 = \sqrt{\delta\phi_1^2 + (\pi -\delta\phi_2)^2}$. Here $\delta \phi_{1}$ ($\delta \phi_{2}$) 
is the difference between azimuthal angle of missing transverse momentum and azimuthal angle of first (second) jet, i.e., $\delta\phi_{1,2}= |\phi_{\ptmiss} - \phi_{J_{1,2}} |$. The $R_1$ and $R_2$ must be greater 
than 0.5 radian. Also $|\phi_{\ptmiss} - \phi_{J} |$ must be greater than 0.3 radian. Also 
we take $|\phi_{\ptmiss} - \phi_{J_2} |$ to be greater than $20^{\circ}$. These cuts are 
very useful to reject QCD backgrounds, where the source of missing energy is jet energy mis-measurements. 

\item The effective mass ($M_T$) of the event must be greater than 500 GeV. Here $M_T$ is defined as 
the scalar sum of  second, third and fourth (if present) jets and $\ptmiss$. 

\end{itemize}

After passing the above mentioned cuts we choose events for further analysis. We take the 
hardest jet and decluster it into two subjets, if the $p_T$ of both subjets are greater than 
 5\% of the parent jet $p_T$ and  these are not too close, i.e. 
$|\delta \eta | + |\delta \phi| > 0.1$. If the declustering is possible, then we take the 
declustered objects and repeat the same procedure on those objects. This procedure 
is terminated if only one calorimeter cell is left. For our work we take a
calorimeter cell of 0.1 in both $\eta$ and $\phi$. If we start with a top 
jet (for hadronic top decay), we will end up 
with three or four subjets. The jet mass of the top jet should lie in between 
150 to 200 GeV. One combination of invariant mass of  the two subjets among 
these 3 or 4 jets must be around W boson mass. Also we demand the cosine of helicity 
angle to be greater than 0.7, where the helicity angle is 
defined as the angle between the top quark direction and one of the decay 
product of W, in the W rest frame. It is also possible to tag one subjet 
as a b jet. However we do not apply b tagging condition in our analysis.
If the $p_T$ of the other sub leading jets are greater than 300 GeV, we apply 
the same algorithm on it and if it satisfies all the criteria,  we declare  
that jet as a top jet. \\

\begin{figure}[t]
\begin{center}
%
\hskip -15pt
{\epsfig{file=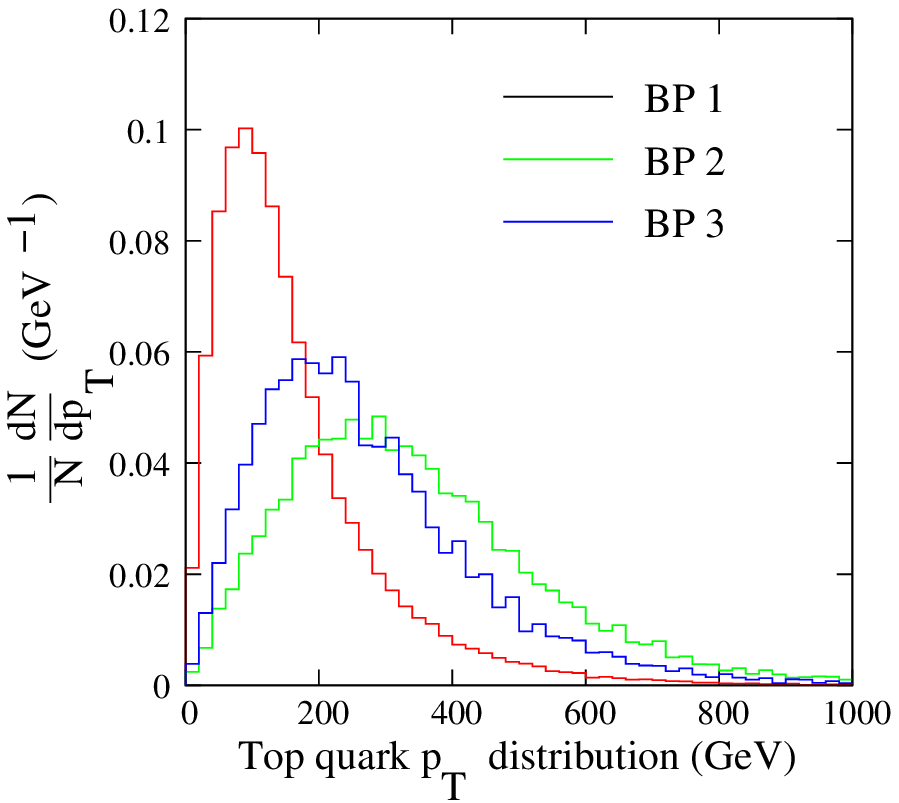,width=8.0 cm,height=8cm,angle=-0}}
\hskip  12pt 
{\epsfig{file=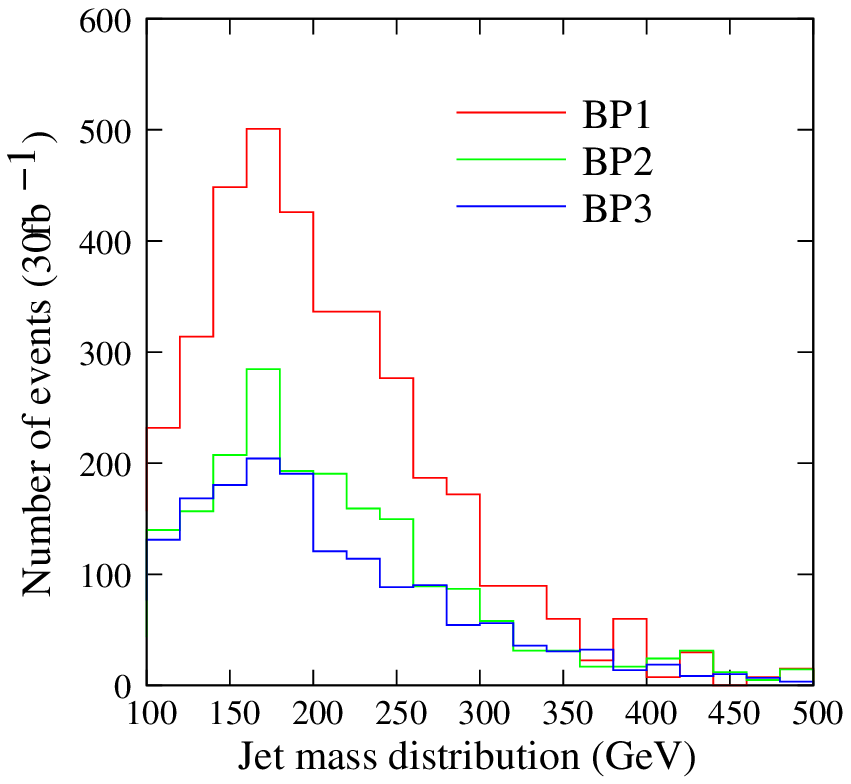,width=8.0cm,height=8cm,angle=-0}}
%
\caption{\sf $p_T$ distribution of the top quark (left) and the jet mass distribution after selection cuts (right).}\label{result}
\end{center}
\end{figure}

In Figure \ref{result} we show the normalized $p_T$ distribution of partonic 
top quarks present in the SUSY cascade for three chosen benchmark points. 
From this plot we have seen that a large fraction of top quark events have 
momenta, greater than 300 GeV. In the right hand side of Figure \ref{result} we 
show the jet mass distribution of events which satisfy all but the 
jet mass cut of the top quark. We observe peaks around top mass in the jet 
mass distributions which verify correctness of the top tagging technique 
used in the analysis. Depending on the parameter space, the peak could be broad 
as it suffers from SUSY contamination in cascade decays. 
QCD, $W/Z$+ jets, di-boson + jets backgrounds are not taken in the analysis, 
though the most significant background, $t \bar{t}$  has been analysed. CMS quotes 
\cite{zerolepton} total backgrounds to be of the order of 250 fb for similar 
type of analysis (with less harder cuts used in our analysis) and QCD contribution
to it is around 100 fb. Here, we have taken much harder cuts on jet $p_T$ 
and missing transverse energy compared to CMS cuts\footnote{We take leading jet $p_T$ to be greater 
than 300 GeV, where they have used 180 GeV cut on it. For comparison see reference \cite{zerolepton}.} and this would 
further reduce the background. One should note that we have used different jet 
formation algorithm with larger jet cone size. Therefore, the size of the SM 
backgrounds can be different from the CMS analysis. However, the probability of 
a QCD jet, mistagged as a top jet is very small and it is $\sim$ 1\% for jet 
$p_T$ 300 GeV \cite{kaplan}. Further, it is possible to add b-tagging criterion 
that will also help to reduce QCD and W backgrounds. \\

In Table \ref{tabsig}, we show the individual contributions coming 
from  $\sstop1\sstopc1$, $\sbot1 \sbotc1$ and $\tilde{g}\tilde{g}$, 
$\tilde{q}\tilde{g}$ productions in the top signal 
for the three benchmark points. We do not show the contributions from $\sstop2$ and $\tilde{b}_2$ separately, 
but these are added to the total contribution. We may recall that the first two generation squarks are heavier than gluino in our benchmark points. These also contribute to the top quark final state through gluino decay. 
The background coming from top quark pair production is not large and it is about 
4-5 fb. We present our result for integrated luminosity of 30 fb$^{-1}$. 
The numbers indicate that even if all backgrounds are considered, the signal would 
be sufficient enough to be observed with 30 fb$^{-1}$ of integrated luminosity at the LHC with 
$\sqrt{s}=14$ TeV. The Table also shows that the gluino contribution is much 
larger than other contributions like stop and sbottom and it is true for all 
benchmark points. As gluino contribution is too large, it may be difficult to 
determine the hierarchy among the third generation squarks and gluino. 
A detailed study should be worked out to see the prospect of such analysis.

\begin{table}[tb]
\begin{center}
\begin{tabular}{||c|c|c|c|c|c||c|} \hline\hline

No.&Benchmark points&$\sstop1\sstopc1$ &$\tilde{b}_1\tilde{b}_1$&$\tilde{g}\tilde{g}$& $\tilde{g}\tilde{q}$&Total \\
   &   & (30 fb$^{-1}$) & (30 fb$^{-1}$ ) & (30 fb$^{-1}$) & (30 fb$^{-1}$) & (30 fb$^{-1}$) \\ 
\hline\hline
1&BP1&     15  &     6  &    142  &     618  &      992   \\
\hline  
2&BP2&      8  &     3  &    110  &     336  &      591    \\
\hline 
3&BP3&      7  &     5  &     42  &     282   &     463    \\
\hline\hline
\end{tabular}
\vspace*{0.0mm}
\caption{\sf Event rates after top tagging for the benchmark points with an integrated luminosity of 30 fb$^{-1}$. 
The $t\bar{t}$ contribution is 132 events assuming same integrated luminosity.}\label{tabsig}
\end{center}
\end{table}

\section{Conclusion}
In this paper we have studied the production of boosted top quarks in the SUSY cascade at the LHC. 
In particular, we have taken a specific SUSY breaking mechanism, cMSSM, although the outcome of 
the study is applicable to any general SUSY breaking scenarios or any other BSM scenarios. 
We systematically discussed different possibilities of top quark final state in the SUSY cascade. 
Top quark identification is difficult in cascade decays due to the combinatorial backgrounds.
Depending on the parameter space, SUSY cascade may have top quarks with high $p_T$. We take advantage of  
this high $p_T$ behaviour of top quark for its identification by using jet substructure algorithm.
This helps to enhance the signal significance for the top final state. It has been vindicated in our 
study that for wide range of SUSY parameter space, this technique is fairly applicable. This study will help 
to understand the third generation squarks and gluino masses and decay modes. This could also be useful 
to probe the regions in SUSY parameter space, where squarks are inaccessible at the LHC but gluino
is in the lighter side. However, we have not done detailed background analysis which would be crucial. 
Also, supersymmetric backgrounds may affect the signal significance and it is very much parameter space 
dependent. One should invoke a more sophisticated analysis to overcome such backgrounds which is beyond the 
scope of our study.

\vskip 20pt

{\bf Acknowledgments:} \\
The authors would like to thank the organisers of LCWS10 (China, 2010), where initial part of this 
work was done. PB would like to thank KIAS Overseas Travel Grant for travel support and BB would 
like to thank India-ILC forum for travel support to attend LCWS10. BB acknowledge discussions with 
Sreerup Raychoudhuri, K Sridhar and M. Guchait.


\begin{thebibliography}{99}
\bibitem{Langacker:2009my}
  P.~Langacker,
  arXiv:0901.0241 [hep-ph].




\bibitem{SUSY}
  M.~Drees, R.~Godbole and P.~Roy,
{\it  Hackensack, USA: World Scientific (2004) 555 p};\\
{For a recent review, see}
  M.~E.~Peskin,
  arXiv:0801.1928 [hep-ph].
\bibitem{Amsler:2008zzb}
  C.~Amsler {\it et al.}  [Particle Data Group],
  Phys.\ Lett.\  B {\bf 667}, 1 (2008).


\bibitem{SUSYSCAN}
  N.~Arkani-Hamed, G.~L.~Kane, J.~Thaler and L.~T.~Wang,
  JHEP {\bf 0608}, 070 (2006);
  D.~Feldman, Z.~Liu and P.~Nath,
  Phys.\ Rev.\ Lett.\  {\bf 99}, 251802 (2007)
  [Erratum-ibid.\  {\bf 100}, 069902 (2008)];
  C.~F.~Berger, J.~S.~Gainer, J.~L.~Hewett and T.~G.~Rizzo,
  JHEP {\bf 0902}, 023 (2009);
  P.~Konar, K.~T.~Matchev, M.~Park and G.~K.~Sarangi,
  Phys.\ Rev.\ Lett.\  {\bf 105}, 221801 (2010).


\bibitem{mSUGRA}
  E.~Cremmer, B.~Julia, J.~Scherk, P.~van Nieuwenhuizen, S.~Ferrara and L.~Girardello,
  Phys.\ Lett.\  B {\bf 79}, 231 (1978);
  E.~Cremmer, B.~Julia, J.~Scherk, S.~Ferrara, L.~Girardello and P.~van Nieuwenhuizen,
  Nucl.\ Phys.\  B {\bf 147}, 105 (1979);
  R.~Barbieri, S.~Ferrara and C.~A.~Savoy,
  Phys.\ Lett.\  B {\bf 119}, 343 (1982);
  A.~H.~Chamseddine, R.~L.~Arnowitt and P.~Nath,
  Phys.\ Rev.\ Lett.\  {\bf 49}, 970 (1982);
  L.~J.~Hall, J.~D.~Lykken and S.~Weinberg,
  Phys.\ Rev.\  D {\bf 27} (1983) 2359;
  P.~Nath, R.~L.~Arnowitt and A.~H.~Chamseddine,
  Nucl.\ Phys.\  B {\bf 227}, 121 (1983);
  N.~Ohta,
  Prog.\ Theor.\ Phys.\  {\bf 70}, 542 (1983).


\bibitem{SUSYSEARCH} {For recent works, see, for example,}
  S.~Bhattacharya, U.~Chattopadhyay, D.~Choudhury, D.~Das and B.~Mukhopadhyaya,
  Phys.\ Rev.\  D {\bf 81}, 075009 (2010);
  J.~Edsjo, E.~Lundstrom, S.~Rydbeck and J.~Sjolin,
  JHEP {\bf 1003}, 054 (2010);
  D.~Feldman, G.~Kane, R.~Lu and B.~D.~Nelson,
  Phys.\ Lett.\  B {\bf 687}, 363 (2010);
  H.~K.~Dreiner, M.~Kramer, J.~M.~Lindert and B.~O'Leary,
  JHEP {\bf 1004}, 109 (2010);
  H.~Baer, V.~Barger, A.~Lessa and X.~Tata,
  JHEP {\bf 1006}, 102 (2010);
  N.~Bhattacharyya, A.~Datta and S.~Poddar,
  Phys.\ Rev.\  D {\bf 82}, 035003 (2010);
  U.~Chattopadhyay, D.~Das, D.~K.~Ghosh and M.~Maity,
  Phys.\ Rev.\  D {\bf 82}, 075013 (2010);
ATLAS Collaboration, ATL-PHYS-PUB-2010-010 (2010);
CMS Collaboration, CMS NOTE 2010/008 (2010);
  B.~Altunkaynak, M.~Holmes, P.~Nath, B.~D.~Nelson and G.~Peim,
  arXiv:1008.3423 [hep-ph];
  B.~C.~Allanach, S.~Grab and H.~E.~Haber,
  arXiv:1010.4261 [hep-ph].
  N.~Chen, D.~Feldman, Z.~Liu, P.~Nath and G.~Peim,
  arXiv:1011.1246 [hep-ph].

\bibitem{zerolepton}
  T.~Yetkin and M.~Spiropulu  [CMS Collaboration],
  Acta Phys.\ Polon.\  B {\bf 38}, 661 (2007);
 M.~Spiropulu,
  Eur.\ Phys.\ J.\  C {\bf 59}, 445 (2009);
J.~B.~G.~da Costa {\it et al.}  [Atlas Collaboration],
  Phys.\ Lett.\  B {\bf 701}, 186 (2011)
  [arXiv:1102.5290 [hep-ex]];
 S.~Chatrchyan {\it et al.}  [CMS Collaboration],
  arXiv:1107.1279 [hep-ex].
 


\bibitem{onelepton}
  Yu.~Pakhotin {\it et al.},
  Acta Phys.\ Polon.\  B {\bf 38}, 653 (2007);
  Yu.~Pakhotin {\it et al.}, CERN-CMS-NOTE-2006-134;
 G.~Aad {\it et al.}  [Atlas Collaboration],
  Phys.\ Rev.\ Lett.\  {\bf 106}, 131802 (2011)
  [arXiv:1102.2357 [hep-ex]];
S.~Chatrchyan {\it et al.}  [CMS Collaboration],
  arXiv:1107.1870 [hep-ex].


\bibitem{stop_sbottom}
  G.~Mahlon and G.~L.~Kane,
  Phys.\ Rev.\  D {\bf 55}, 2779 (1997);
  W.~Beenakker, M.~Kramer, T.~Plehn, M.~Spira and P.~M.~Zerwas,
  Nucl.\ Phys.\  B {\bf 515}, 3 (1998);
  R.~Demina, J.~D.~Lykken, K.~T.~Matchev and A.~Nomerotski,
  Phys.\ Rev.\  D {\bf 62}, 035011 (2000);
  A.~Dedes and H.~K.~Dreiner,
  JHEP {\bf 0106}, 006 (2001);
  G.~Bozzi, B.~Fuks and M.~Klasen,
  Phys.\ Rev.\  D {\bf 72}, 035016 (2005);
  W.~Hollik, M.~Kollar and M.~K.~Trenkel,
  JHEP {\bf 0802}, 018 (2008);
  M.~Beccaria, G.~Macorini, L.~Panizzi, F.~M.~Renard and C.~Verzegnassi,
  Int.\ J.\ Mod.\ Phys.\  A {\bf 23}, 4779 (2008)
 S.~Bornhauser, M.~Drees, S.~Grab {\it et al.},
[arXiv:1011.5508 [hep-ph]].

\bibitem{Agashe:2006hk}
  K.~Agashe, A.~Belyaev, T.~Krupovnickas {\it et al.},
  Phys.\ Rev.\  {\bf D77}, 015003 (2008).
  [hep-ph/0612015].

\bibitem{boostedtop}
  M.~M.~Nojiri and M.~Takeuchi,
  JHEP {\bf 0810}, 025 (2008);
  J.~Thaler and L.~T.~Wang,
  JHEP {\bf 0807}, 092 (2008);
  L.~G.~Almeida, S.~J.~Lee, G.~Perez, I.~Sung and J.~Virzi,
  Phys.\ Rev.\  D {\bf 79}, 074012 (2009);
  D.~Krohn, J.~Shelton and L.~T.~Wang,
  JHEP {\bf 1007}, 041 (2010);
  B.~Bhattacherjee, M.~Guchait, S.~Raychaudhuri and K.~Sridhar,
  Phys.\ Rev.\  D {\bf 82}, 055006 (2010);
  K.~Rehermann and B.~Tweedie,
  arXiv:1007.2221 [hep-ph].


\bibitem{kaplan}
  D.~E.~Kaplan, K.~Rehermann, M.~D.~Schwartz and B.~Tweedie,
  Phys.\ Rev.\ Lett.\  {\bf 101}, 142001 (2008).


\bibitem{boostedhiggs}
  J.~M.~Butterworth, A.~R.~Davison, M.~Rubin and G.~P.~Salam,
  Phys.\ Rev.\ Lett.\  {\bf 100}, 242001 (2008);
  T.~Plehn, G.~P.~Salam, M.~Spannowsky,
  Phys.\ Rev.\ Lett.\  {\bf 104}, 111801 (2010);
  G.~D.~Kribs, A.~Martin, T.~S.~Roy {\it et al.},
  Phys.\ Rev.\  {\bf D81}, 111501 (2010);
  D.~E.~Soper, M.~Spannowsky,
  JHEP {\bf 1008}, 029 (2010);
  C.~-R.~Chen, M.~M.~Nojiri, W.~Sreethawong,
  JHEP {\bf 1011}, 012 (2010);
  A.~Falkowski, D.~Krohn, L.~T.~Wang, J.~Shelton and A.~Thalapillil,
  arXiv:1006.1650 [hep-ph];
  G.~D.~Kribs, A.~Martin, T.~S.~Roy and M.~Spannowsky,
  arXiv:1006.1656 [hep-ph].
  C.~Hackstein and M.~Spannowsky,
  arXiv:1008.2202 [hep-ph];
  J.~-H.~Kim,
[arXiv:1011.1493 [hep-ph]];
  B.~Bellazzini, C.~Csaki, J.~Hubisz and J.~Shao,
  arXiv:1012.1316 [hep-ph];
\bibitem{boostedw}
  J.~M.~Butterworth, J.~R.~Ellis, A.~R.~Raklev,
  JHEP {\bf 0705}, 033 (2007);
  L.~G.~Almeida, S.~J.~Lee, G.~Perez {\it et al.},
  Phys.\ Rev.\  {\bf D79}, 074017 (2009);
  Y.~Cui, Z.~Han and M.~D.~Schwartz,
  arXiv:1012.2077 [hep-ph];
  J.~Thaler and K.~Van Tilburg,
  arXiv:1011.2268 [hep-ph].



\bibitem{Plehn}
  T.~Plehn, M.~Spannowsky, M.~Takeuchi and D.~Zerwas,
  JHEP {\bf 1010}, 078 (2010).




\bibitem{thirdgeneration}
  S.~P.~Martin, P.~Ramond,
  Phys.\ Rev.\  {\bf D48 } (1993)  5365-5375;
  S.~P.~Martin,
  In *Kane, G.L. (ed.): Perspectives on supersymmetry* 1-98.
  [hep-ph/9709356].

\bibitem{Bartl:1998xk}
  A.~Bartl {\it et al.},
  Phys.\ Lett.\  B {\bf 435}, 118 (1998)
  [arXiv:hep-ph/9804265].

\bibitem{cpc_cascade}
  A.~Datta, M.~Guchait, K.~K.~Jeong,
  Int.\ J.\ Mod.\ Phys.\  {\bf A14}, 2239-2256 (1999);
  A.~Djouadi, Y.~Mambrini,
  Phys.\ Lett.\  {\bf B493}, 120-126 (2000);
 A.~Djouadi, M.~Guchait, Y.~Mambrini,
  Phys.\ Rev.\  {\bf D64}, 095014 (2001);
  A.~Djouadi, Y.~Mambrini,
  Phys.\ Rev.\  {\bf D63}, 115005 (2001);
  A.~Belyaev, J.~R.~Ellis, S.~Lola,
  Phys.\ Lett.\  {\bf B484}, 79-86 (2000);
  A.~Datta, A.~Djouadi, M.~Guchait {\it et al.},
  Phys.\ Rev.\  {\bf D65}, 015007 (2002);
  M.~Graesser, J.~Shelton,
  JHEP {\bf 0906}, 039 (2009).



\bibitem{cpv_cascade}
  P.~Langacker, G.~Paz, L.~-T.~Wang {\it et al.},
  JHEP {\bf 0707}, 055 (2007);
  A.~Bartl, E.~Christova, K.~Hohenwarter-Sodek {\it et al.},
  JHEP {\bf 0611}, 076 (2006);
  J.~Ellis, F.~Moortgat, G.~Moortgat-Pick {\it et al.},
  Eur.\ Phys.\ J.\  {\bf C60}, 633-651 (2009);
  G.~Moortgat-Pick, K.~Rolbiecki, J.~Tattersall {\it et al.},
  JHEP {\bf 1001}, 004 (2010);
  G.~Moortgat-Pick, K.~Rolbiecki, J.~Tattersall {\it et al.},
  AIP Conf.\ Proc.\  {\bf 1200}, 337-340 (2010);
  P.~Bandyopadhyay,
[arXiv:1008.3339 [hep-ph]].
\bibitem{stoploop}
  K.~i.~Hikasa and M.~Kobayashi,
  Phys.\ Rev.\  D {\bf 36}, 724 (1987).

\bibitem{Porod:1996at}
  W.~Porod and T.~Wohrmann,
  Phys.\ Rev.\  D {\bf 55}, 2907 (1997)
  [Erratum-ibid.\  D {\bf 67}, 059902 (2003)]
  [arXiv:hep-ph/9608472].


\bibitem{stop4body} 
  C.~Boehm, A.~Djouadi, Y.~Mambrini,
  Phys.\ Rev.\  {\bf D61}, 095006 (2000);
  S.~P.~Das, A.~Datta and M.~Guchait,
  Phys.\ Rev.\  D {\bf 65}, 095006 (2002).

\bibitem{SUSY-HIT}
  A.~Djouadi, M.~M.~Muhlleitner and M.~Spira,
  Acta Phys.\ Polon.\  B {\bf 38}, 635 (2007)
  [arXiv:hep-ph/0609292].


\bibitem{gluinodecay}
 H.~Baer, V.~D.~Barger, D.~Karatas and X.~Tata,
  Phys.\ Rev.\  D {\bf 36}, 96 (1987);
 R.~M.~Barnett, J.~F.~Gunion and H.~E.~Haber,
  Phys.\ Rev.\  D {\bf 37}, 1892 (1988);
 A.~Bartl, W.~Majerotto, B.~Mosslacher, N.~Oshimo and S.~Stippel,
  Phys.\ Rev.\  D {\bf 43}, 2214 (1991);
 J.~Hisano, K.~Kawagoe, R.~Kitano {\it et al.},
 Phys.\ Rev.\  {\bf D66}, 115004 (2002);
 P.~Gambino, G.~F.~Giudice, P.~Slavich,
 Nucl.\ Phys.\  {\bf B726}, 35-52 (2005);
  B.~S.~Acharya, P.~Grajek, G.~L.~Kane, E.~Kuflik, K.~Suruliz and L.~T.~Wang,
  V.~Barger, W.~Y.~Keung and B.~Yencho,
  Phys.\ Lett.\  B {\bf 687}, 70 (2010);
  G.~F.~Giudice, T.~Han, K.~Wang and L.~T.~Wang,
  Phys.\ Rev.\  D {\bf 81}, 115011 (2010);
 A.~Bartl, K.~Hidaka, K.~Hohenwarter-Sodek, T.~Kernreiter, W.~Majerotto and W.~Porod,
  Phys.\ Lett.\  B {\bf 679}, 260 (2009)
  [arXiv:0905.0132 [hep-ph]];
  A.~Bartl, H.~Eberl, E.~Ginina, B.~Herrmann, K.~Hidaka, W.~Majerotto and W.~Porod,
  arXiv:1107.2775 [hep-ph].

\bibitem{Bartl:1994bu}
  A.~Bartl, W.~Majerotto and W.~Porod,
  Z.\ Phys.\  C {\bf 64}, 499 (1994)
  [Erratum-ibid.\  C {\bf 68}, 518 (1995)].



\bibitem{focuspoint}
  K.~L.~Chan, U.~Chattopadhyay and P.~Nath,
  Phys.\ Rev.\  D {\bf 58}, 096004 (1998);
  J.~L.~Feng, K.~T.~Matchev and T.~Moroi,
  Phys.\ Rev.\  D {\bf 61}, 075005 (2000);
  U.~Chattopadhyay, A.~Datta, A.~Datta, A.~Datta and D.~P.~Roy,
  Phys.\ Lett.\  B {\bf 493}, 127 (2000);
  J.~L.~Feng and F.~Wilczek,
  Phys.\ Lett.\  B {\bf 631}, 170 (2005);
 H.~Baer, V.~Barger, G.~Shaughnessy {\it et al.},
 Phys.\ Rev.\  {\bf D75}, 095010 (2007);
  S.~P.~Das, A.~Datta, M.~Guchait, M.~Maity and S.~Mukherjee,
  Eur.\ Phys.\ J.\  C {\bf 54}, 645 (2008);
  N.~Desai and B.~Mukhopadhyaya,
  Phys.\ Rev.\  D {\bf 80}, 055019 (2009).


\bibitem{bhattacherjee}
B.~Bhattacherjee, A.~Dighe, D.~Ghosh and S.~Raychaudhuri,
  Phys.\ Rev.\  D {\bf 83}, 094026 (2011)
  [arXiv:1012.1052 [hep-ph]].

\bibitem{superiso}
 F.~Mahmoudi,
  Comput.\ Phys.\ Commun.\  {\bf 178}, 745 (2008)
  [arXiv:0710.2067 [hep-ph]];
F.~Mahmoudi,
  Comput.\ Phys.\ Commun.\  {\bf 180}, 1579 (2009)
  [arXiv:0808.3144 [hep-ph]].

\bibitem{Heinemeyer}
S.~Heinemeyer,
  Int.\ J.\ Mod.\ Phys.\  A {\bf 21}, 2659 (2006)
  [arXiv:hep-ph/0407244].

\bibitem{micromegas}
G.~Belanger, F.~Boudjema, A.~Pukhov and A.~Semenov,
  Comput.\ Phys.\ Commun.\  {\bf 149}, 103 (2002)
  [arXiv:hep-ph/0112278];
G.~Belanger, F.~Boudjema, A.~Pukhov and A.~Semenov,
  Comput.\ Phys.\ Commun.\  {\bf 176}, 367 (2007)
  [arXiv:hep-ph/0607059].

\bibitem{WMAP}
  E.~Komatsu {\it et al.}  [WMAP Collaboration],
  Astrophys.\ J.\ Suppl.\  {\bf 192}, 18 (2011)
  [arXiv:1001.4538 [astro-ph.CO]].
\bibitem{cambridge_aachen}
 Y.~L.~Dokshitzer, G.~D.~Leder, S.~Moretti and B.~R.~Webber,
 JHEP {\bf 9708}, 001 (1997);
 M.~Wobisch, T.~Wengler,
 [hep-ph/9907280].


\bibitem{PYTHIA}
  T.~Sjostrand, S.~Mrenna and P.~Z.~Skands,
  JHEP {\bf 0605}, 026 (2006)
  [arXiv:hep-ph/0603175].

\bibitem{FASTJET}
  M.~Cacciari and G.~P.~Salam,
  Phys.\ Lett.\  B {\bf 641}, 57 (2006)
  [arXiv:hep-ph/0512210].




\end{thebibliography}
\end{document}